\documentclass[11pt,twoside]{article}

\usepackage{asp2006}
\usepackage{epsf}
\usepackage{psfig}
\usepackage{lscape}

\begin{document}
\title{The PennState/Toru\'n Center for Astronomy  Search for Planets Around Evolved Stars.\\ Basic parameters of a sample of evolved stars}   
\begin{quote}
Pawe\l{} Zieli\'nski$^1$ and Andrzej Niedzielski$^{1,2}$

{\itshape $^1$Torun Centre for Astronomy, Nicolaus Copernicus University,\\
ul. Gagarina 11, 87-100 Torun, Poland}

{\itshape $^2$Department of Astronomy and Astrophysics, Pennsylvania State
University, 525 Davey Laboratory, University Park, PA 16802}
\end{quote}

\begin{abstract} 
The objective of the PSU/TCfA Search for Planets Around Evolved Stars is to study evolution of planetary systems in the stellar evolution timescale. For such an analysis precise physical parameters of the hosts of the  planetary systems are essential. In this paper we present an attempt to obtain basic physical parameters for a sample of evolved stars observed within our survey with the High Resolution Spectrograph of the Hobby-Eberly Telescope. 
\end{abstract}

\section*{Introduction}
Proper interpretation of results obtained from radial velocity (RV) survey of GK giants requires detailed knowledge of their physical parameters. Since GK-giants are known to exhibit  various types of intrinsic variability induced by pulsations and/or surface inhomogeneity connected with rotation these effects  have to be ruled out before substellar companion interpretation can be considered.  Effective temperatures and gravitational accelerations are needed to obtain estimates of stellar radii and together with the rotational periods address the influence of possible starspots on results. Additional knowledge of metallicities allows for stellar masses estimates that are crucial for interpretation of the observed RV periodical variations and determination of companion's masses. 

The observational material and data reduction are described in Niedzielski \& Wolszczan (this volume).

\section*{Results and Conclusions}

The atmospheric parameters of program stars were obtained with the purely spectroscopic method \citep{Takeda02,Takeda05a}  based on analysis of Fe I and Fe II lines and  assumption of LTE. 

To test  reliability of our determinations we used the procedure to obtain parameters of 8 stars of \cite{2006ApJ...646..505B}.  Comparison of results shows that $T_{eff}$ agrees within $49 K$, $log g$ within $0.11 dex$ and $[Fe/H]$ within $0.11 dex$. 
These values are comparable to  our  intrinsic error estimates, except for metallicity, which is systematically higher than literature values.
Typical intrinsic uncertainties of these parameters are $\sigma{T_{eff}} = 29 K$, $\sigma{log g} = 0.09$, $\sigma{v_t} = 0.13 km s^{-1}$ and $\sigma{[Fe/H]} = 0.06$. 
In Table 1 we present results of our determinations of atmospheric parameters for 16 stars, targets of our survey.  The metalicities presented in Table 1 are corrected  for the systematic effect.
 Stellar masses were obtained  by comparing the position of stars on the HR diagram with the theoretical evolutionary tracks of \citet{Girardi00} and \citet{Salasnich00} for given metallicity. Stellar radii were determined using the calibration given in \citet{Alonso00}.
The uncertainties of stellar masses obtained by comparison with theoretical evolutionary tracks depend on the accuracy of the position of a star on HR diagram and metalicity. For stars for which the parallax is precise enough the metallicity may introduce significant uncertainty in mass. We assume that for an average red giant with known parallax the mass may be estimated within $0.3 M_\odot$.

\begin{table}[!ht]
\caption{Determined parameters for program stars}
\smallskip
\begin{center}
\begin{tabular}{l|c|c|c|c|c|c|c}
\tableline
\tableline
\noalign{\smallskip}
Name & $T_{eff}$ & $log g$ & $v_t$ & $[Fe/H]$ & $M_v$ & $M/M_o$ & $R/R_o$\\
& $(K)$ & & $(km/s)$ & & $(mag)$ & &\\
\noalign{\smallskip}
\tableline
\noalign{\smallskip}
HD 17028	 & 4246 & 2.43 & 1.52 & 0.11 & -2.89 & 5.5 & 27.4\\
HD 17092   & 4649 & 3.00 & 1.27 & 0.22 & 0.80 & 2.3 & 10.9\\
HD 77819   & 5007 & 2.92 & 1.19 & -0.11 & 1.01 & 2.5 & 9.4\\
HD 95296	 & 4603 & 2.23 & 1.36 & -0.08 & -0.28 & 2.7 & 10.9\\
HD 96127	 & 4319 & 2.65 & 1.23 & 0.35 & -2.46 & 5.0 & 22.1\\
HD 102103	 & 4565 & 2.54 & 1.44 & -0.04 & -0.63 & 3.0 & 12.8\\
HD 102272  & 4908 & 3.07 & 1.30 & -0.26 & 1.43 & 1.7 & 10.1\\
HD 109402  & 5036 & 4.03 & 0.80 & -0.11 & 3.34 & 1.3 & 8.6\\
HD 112914  & 4975 & 4.86 & 0.01 & -0.25 & 5.22 & 0.8 & -\\
HD 134901  & 5427 & 3.91 & 0.86 & -0.22 & 3.09 & 1.3 & 3.4\\
HD 160723  & 4775 & 2.67 & 1.29 & -0.30 & 0.80 & 2.2 & 10.3\\
HD 257498  & 4773 & 2.94 & 1.33 & -0.19 & 5.85 & 0.8 & 10.3\\
BD +03 2562& 4365 & 2.95 & 1.29 & -0.12 & 0.40 & 1.4 & 22.1\\
BD +20 2457& 4352 & 2.19 & 1.64 & -0.65 & 0.40 & 0.8 & 22.1\\
BD +57 144 & 5010 & 3.09 & 1.28 & -0.08 & -0.22 & 3.3 & 9.4\\
SAO 112133 & 4993 & 3.56 & 1.50 & 0.18 & 0.80 & 2.1 & 9.4\\
\noalign{\smallskip}
\tableline
\end{tabular}
\end{center}
\end{table}

\acknowledgements  
We thank Y. Takeda for making his code available for us. The project is  supported in part by the Polish Ministry of Science and Higher Education grant 1P03D 007 30.  PZ was supported by Polish Ministry of Science and Higher Education grant SPB 104/E-337/6. 


\end{document}